\documentclass[prl,twocolumn,tightlines,showpacs,preprintnumbers,aps,floatfix,showpacs,amsmath,amssymb]{revtex4}
\usepackage{epsfig,amsmath,amssymb,bm,epsf,graphics}
\usepackage{dcolumn}

\newcommand{\be}{\begin{equation}}
\newcommand{\ee}{\end{equation}}
\newcommand{\ben}{\begin{eqnarray}}
\newcommand{\een}{\end{eqnarray}}
\newcommand{\bF}{\begin{figure}}
\newcommand{\eF}{\end{figure}}
\newcommand{\dg}{\dagger}

\begin{document}

\title{Information transfer through a one-atom micromaser}
\author{Animesh Datta}
\affiliation{Department of Physics and Astronomy, University
of New Mexico, Albuquerque, New Mexico 87131-1156, USA}
\email{animesh@unm.edu}
\author{Biplab Ghosh}
\email{biplab@bose.res.in}
\author{A. S. Majumdar}
\email{archan@bose.res.in}
\author{N. Nayak}
\email{nayak@bose.res.in}
\affiliation{S. N. Bose National Centre for Basic Sciences, Block JD, Salt Lake, Kolkata 700 098, INDIA}

\date{\today}

\begin{abstract}
We consider a realistic model for the  one-atom micromaser consisting of a 
cavity maintained in 
a steady state by the streaming of  two-level Rydberg 
atoms passing one at a time
through it. We show that it is possible to monitor the robust entanglement 
generated between two successive experimental atoms passing through the cavity 
by the control decoherence parameters. We calculate 
the entanglement of formation of the joint two-atom state as a function of the
micromaser pump parameter. We find that this
is in direct correspondence with the difference of the Shannon entropy
of the cavity photons before and after the passage of the atoms for a 
reasonable range of dissipation parameters. It is thus
possible to demonstrate information transfer between the cavity
and the atoms through this set-up.
\end{abstract}



\maketitle

The generation of quantum entanglement in atomic systems is being vigorously
pursued in recent years. The primary motivation for this upsurge of interest
is to test the applicability of the  ongoing conceptual developments
in quantum information theory and through them the implementation of current
quantum communication and computation protocols\cite{nielsen}.
Several schemes have been proposed recently to engineer the entanglement
of two\cite{cabrillo} or more 
atoms\cite{duan,solano}. Many of these proposals are for generating 
entanglement
in a probabilistic manner.  Since a large number of these proposals 
rely on the trapping or slow passage of cold atoms
through optical cavities\cite{raimond}, the efficient control of cavity 
leakage and atomic dissipation
is a major concern\cite{tregenna}. The value of the atom-cavity 
coupling $g$ is very close to the values
of photonic and atomic decay rates $\kappa$ and $\Gamma$, respectively, in
the parameter ranges operated by optical cavities. Thus decoherence effects
are significant even in the time $O(1/g)$ needed for perceptible entanglement.

The micromaser, described below, is appreciated as a practical device 
for processing quantum
information. The formation of atom-photon entanglement and the subsequent
generation of correlations between spatially 
separated atoms has been shown using the micromaser. The nonlocal correlations 
developed in this fashion between two or more atoms can be used to test the 
violation of Bell-type inequalities\cite{hagley,majumdar,masiak,englert}.
Since for Rydberg atoms tuned with microwave frequencies, $\kappa \ll g$,
and $\Gamma$ is negligible, decoherence does not crucially affect the
individual single atom dynamics. However, dissipation does build up over
the passage of a number of atoms through the micromaser, and is revealed
in the photonic statistics of the steady-state cavity field, as was 
discussed in Ref. \cite{majumdar}. The entanglement
between a pair of atoms pumped at the same time through a micromaser has
been analysed in Ref. \cite{masiak}. It is rather difficult to practically 
realize such a set-up though.  The genuine one-atom 
micromaser, on the other hand, can be
operated over a reasonably large region of parameter space, and is thus
a feasible device\cite{rempe} for generating entanglement between two
or more atoms.
Recently, Englert et al\cite{englert} have shown  using a 
non-separability criterion, the generation of entanglement between two atoms 
that pass through a one-atom
micromaser one after the other, in immediate succession. 

In this Letter we propose a scheme to measure the entanglement generated 
between two successive atoms that stream through a real one-atom micromaser
in such a manner that their flights through
the cavity do not overlap. There is always a time gap between one atom
leaving the cavity and the next atom entering the cavity. We show that
successive atoms that emerge out of the cavity are entangled. This scheme
does not require the spatial overlap between the two atoms at any stage.
In the theory for the micromaser used by us \cite{majumdar,nayak}, the 
interaction of the cavity with its reservoir is taken into account at all 
times. We indeed compute the effect of photon leakage on the entanglement
measure. Such a model was earlier used by us to show the violation of
a Bell-type inequality between two spatially non-overlapping atoms correlated
via atom-photon interactions. This strongly suggests entanglement of the
successive atoms that pass through the cavity.
The generation of nonlocal correlations between the two atomic states
emerging from the cavity can in general be understood using the Horodecki
theorem\cite{horodecki}. 

Since the joint state of the two atoms emanating from the cavity is not a
pure state, we quantify the entanglement  using the well known 
measure appropriate for mixed states, i.e., the  entanglement
of formation\cite{hill}. Information is transferred from the cavity to
the atoms in order to build up entanglement. The amount of information
transferred is expected to depend on the available information content of
the cavity. The variance of the photon number distribution of the cavity
is an indicator of the information content of the cavity field, and we
compute the variance over a range of atom-photon interaction times. The
total information inside the cavity is however measured by its Shannon
entropy\cite{shannon} which has contributions from higher moments of the
photon statistics as well.
We therefore calculate the Shannon entropy
of the photon distribution function in the cavity before and after the
passage of the two atoms. The difference in the Shannon entropy is seen
to be in remarkable correspondence with the entanglement of formation 
of the atoms up to a reasonably long atom-photon interaction time.
We can thus claim that this scheme provides a concrete quantitative example
of information transfer between the microwave cavity and the two atoms
in a realistic set-up.
 
We begin with a description of the experimental scenario.
A two-level atom  initially in its upper excited state $|e\rangle$ traverses
a high-Q single mode cavity.
The cavity is in a steady state built up by the passage of a large number of 
atoms, but only one at a time, with a fixed pump parameter and atom-photon
interaction time, and is tuned to a single mode resonant with the transition 
$|e\rangle \to |g\rangle$. The emerging
single-atom wavefunction is a superposition of the upper $|e\rangle$ and
lower $|g\rangle$ state  and it 
leaves an imprint on the photonic wavefunction
in the cavity. During this process, cavity leakage takes place, and is
taken into account. Next, a second experimental atom, prepared also in
state $|e\rangle$, encounters the cavity
photons whose state has been now modified by the passage of the first atom.
This atom too emerges in a superposition with either of the above
two outcomes possible. Although there is no direct interaction between the
atoms, correlations develop between their states mediated by the cavity
photons\cite{hagley,majumdar}. The effects of dissipation on these correlations
can be computed\cite{majumdar}. After emanating from the cavity, 
each of the
atoms encounters a $\pi/2$ pulse through an electromagnetic field whose
phase may be varied for different atoms. The effect of the $\pi/2$ pulse
is to transform the state ${1\over \sqrt{2}}(|e\rangle + |g\rangle)$ to $|g\rangle$ and the
other one to $|e\rangle$. These resultant states may now be detected, and the
corresponding outcomes used to signify the states emanating from the cavity.

The micromaser model has been described in details in Refs 
\cite{majumdar,nayak}. Here we merely outline its essential features.
The cavity is pumped to its steady state by a Poissonian stream
of atoms passing through it one at a time, with the time of flight through
the cavity $t$
being same for every atom.
The dynamics of these individual flights are governed by the evolution 
equation with three kinds
of interactions given by
\be
\dot{\rho} = \dot{\rho}\vert_{atom-reservoir} + \dot{\rho}\vert_{field-reservoir}
+ \dot{\rho}\vert_{atom-field}
\ee
where the strength of the three couplings are given by the parameters $\Gamma$,
(the atomic dissipation constant)
$\kappa$ (the cavity leakage constant) and $g$ (the atom-field interaction
constant, or $gt$ the Rabi angle) and the individual expressions provided 
in Refs \cite{majumdar,nayak}. Obviously, $\Gamma=0=g$ describes the dynamics
of the cavity when there is no atom inside it. The finite temperature of the
cavity is represented by the average thermal photons $n_{{\rm th}}$, obtained
from B-E statistics. The density matrix of the steady state
cavity field $\rho_f^{(ss)}$ can be obtained by solving the above equation and
tracing over the reservoir and atomic variables.
The photon distribution function  is then given  by
\be
P_n^{(ss)} = \langle n|\rho_f^{(ss)}|n\rangle
\ee
in the photon number ($n$) representation,
the expression for which in this model of
the micromaser has been derived in Ref. \cite{nayak}.

The joint state of the two
experimental atoms emanating successively from the cavity is not separable,
and is represented by a mixed density operator  
$\rho_a$ which is obtained by tracing over the field variables. $\rho_a$ can be
written as
\be
\rho_a = \left(\begin{array}{ccccc}
                         \alpha_1 & 0  & 0 & 0  \\
                         0 & \alpha_3 & \alpha_4 {\rm e}^{i\theta} & 0  \\
                         0 & \alpha_4 {\rm e}^{-i\theta}  & \alpha_2 & 0  \\
                         0 & 0  & 0  & \alpha_5 \end{array}\right)
\label{density}
\ee
where $\theta$ is the phase difference between the two atoms introduced by the
$\pi/2$ pulses, and the matrix elements are given by
\ben
\alpha_1 = \mathrm {Tr_f}\biggl({\cal A}{\cal A}\rho_f^{(ss)}
{\cal A}^{\dg}{\cal A}^{\dg}\biggr), \\
\alpha_2 = \mathrm {Tr_f}\biggl({\cal A}{\cal D}\rho_f^{(ss)}
{\cal D}^{\dg}{\cal A}^{\dg}\biggr), \\ 
\alpha_3 = \mathrm {Tr_f}\biggl({\cal D}{\cal A}\rho_f^{(ss)}
{\cal A}^{\dg}{\cal D}^{\dg}\biggr), \\
\alpha_4 = \mathrm {Tr_f}\biggl({\cal D}{\cal A}\rho_f^{(ss)}
{\cal D}^{\dg}{\cal A}^{\dg}\biggr), \\
\alpha_5 = \mathrm {Tr_f}\biggl({\cal D}{\cal D}\rho_f^{(ss)}
{\cal D}^{\dg}{\cal D}^{\dg}\biggr)
\een
The field operators ${\cal A}$ and ${\cal D}$ are defined as
\ben
{\cal A} = {\rm cos}(gt\sqrt{a^{\dg}a + 1})  \\
{\cal D} = -ia^{\dg}{{\rm sin}(gt\sqrt{a^{\dg}a + 1}) \over
\sqrt{a^{\dg}a + 1}}
\een
where $a$($a^{\dg}$) is the usual photon annihilation (creation) operator.

The entanglement of formation  $E_F$ of the two-atom system can be expressed 
as \cite{hill}
\begin{subequations}
\ben
E_F(\rho_a)& = & h\left( \frac{1}{2}[1 + \sqrt{1-C(\rho_a)^2}] \right),
\label{eqn:Wootters} \\
h(x) & \equiv & -x \log_2 x - (1-x) \log_2(1-x),
\label{eqn:entropy}
\een
where $C(\rho_a)$, the {\em concurrence} of the state $\rho_a$, is defined as
\ben
        C(\rho_a) \equiv \max \{0,\sqrt{\lambda_1}-\sqrt{\lambda_2}-
 \sqrt{\lambda_3}-\sqrt{\lambda_4}\},
\een
\end{subequations}
in which 
$\lambda_1,\ldots,\lambda_4$ are the eigenvalues of the matrix 
$\rho_a (\sigma_y\otimes\sigma_y)\rho_a^{*}(\sigma_y\otimes\sigma_y)$ in 
decreasing order and $\sigma_y$ is the Pauli spin matrix. $E_F(\rho_a)$, 
$C(\rho_a)$, and the {\it tangle\/} $\tau(\rho_a)\equiv C(\rho_a)^2$ are equivalent 
measures of entanglement, inasmuch as they are monotonic functions of one 
another.

For the state $\rho_a$ in  Eq. (\ref{density}), the four eigenvalues of the
matrix $\rho_a (\sigma_y\otimes\sigma_y)\rho_a^{*}(\sigma_y\otimes\sigma_y)$
are given by $(\alpha_4 + \sqrt{\alpha_2\alpha_3})^2$, $(\alpha_4 - \sqrt{\alpha_2\alpha_3})^2$, $\alpha_1\alpha_5$, and $\alpha_1\alpha_5$ respectively.
We compute numerically the values of $E_F(\rho_a)$ for a range of micromaser
parameters. In Fig.1 we plot the Entanglement of formation $E_F$ of the 
two-atom state versus the dimensionless cavity dissipation parameter 
$\kappa/g$.
The values of the other parameters, i.e., $N$ (the micromaser pump rate 
denoting the number
of atoms that pass through the cavity in a photon lifetime to maintain
its steady state), $gt$ (the Rabi angle), $n_{th}$ (the number density
of thermal photons) are displayed in the figure caption. It is clearly
seen that cavity dissipation leads to the loss of information transfer
to the joint state of the two atom system. It may be further noted that
the interaction times in the Curves II-IV correspond to the so-called
``trapped states'' conditions\cite{rajagop}. In such a situation, it is
expected that the cavity field would evolve to a Fock state or a set
of discrete states satisfying the condition ${\mathrm sin}(gt\sqrt{n_i+1})=0$.
Naturally, the correlation between different field states $|n_i>$ is not
significant, and further deteriorates with increasing cavity leakage. This
is reflected in Curves II-IV where the fall of entanglement is much faster
than that in the situation where $gt$ does not satisfy ``trapped state''
conditions (Curve I). Note that the nonvanishing values of $E_F$ in 
Curves II-IV suggest that the field does not evolve to an exact Fock
state of the photon field\cite{nayak}. 
\begin{figure}
\begin{center}
\centerline{\epsfig{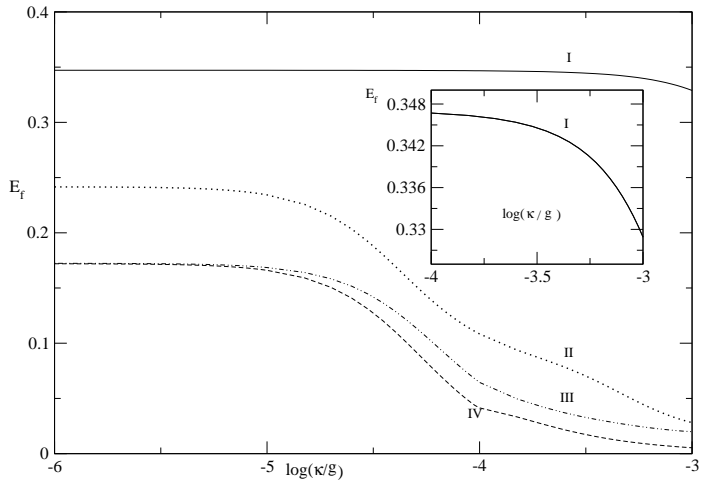}}
\caption{The entanglement of formation $E_F$ plotted with respect 
to the dimensionaless cavity dissipation parameter $\kappa/g$ with
 $N=100$ and $n_{\mathrm th}=0.01$ for different values of the Rabi angle
$gt$. It is seen that $E_F$ falls down with increasing cavity leakage
$\kappa$ for all values of $gt$, including curve I ($gt = \pi/10$) which
is also displayed in the inset for clarity. The other values of $gt$
in curves II ($gt=\pi/2$), III ($gt=3\pi/2$) and IV ($gt=\pi$) each 
corresponds to trapped state conditions.}
\end{center}
\end{figure}

The steady-state photon statistics of the cavity field\cite{majumdar,nayak} 
depending on the cavity parameters
are reflected in the entanglement properties of the emerging atoms. As an
illustration of this we compute the variance of the photon number distribution
$v = \sqrt{[<n^2>-<n>^2]/<n>}$ as function of the pump parameter  
$D=gt\sqrt{N}$. The entanglement of formation $E_F$ and the photon number
variance $(v/10)$ are both plotted versus $D$ in Fig.2. The peaks in $v$ are
related to the thresholds of micromaser operation (these chracteristics of
the micromaser are discussed in details in Refs.\cite{rempe,nayak}), and
correspond to increased uncertainty or information content in the photon
distribution function. It is hence not surprising that maximum information
transfer to the atoms takes place around these parameter values as
signified by the peaks of $E_F$. The functional relation of $E_F$ with
$D$ is reminiscent of the behaviour of the Bell sum \cite{majumdar}
in the same model. It can be shown using the Horodecki theorem\cite{horodecki}
that a Bell-type inequality will be violated in this set-up if and only if 
${\mathcal M}(\rho_a)>1$ where ${\mathcal M}(\rho_a)$ is defined as 
the sum of the two largest eigenvalues of the $3 \times 3$ matrix 
$T^{\dagger}T$
with the elements $T_{ij}= \mathrm{Tr}(\rho_a\sigma_i\otimes\sigma_j)$.  Here
the three eigenvalues of the matrix $T^{\dagger}T$ are given by
$(2\alpha_4)^2$, $(2\alpha_4)^2$, and $(\alpha_1 - \alpha_2 - \alpha_3 +\alpha_5)^2$ respectively. It is possible to confirm numerically that violation of
a Bell-type inequality will occur for a range of parameter values.
  
The role of photon statistics on information transfer is further revealed
by the computation of  the Shannon entropy which for  the steady-state cavity 
is defined as
\be
S(\rho_f^{(ss)}) = -\sum_{n=1}^{\infty}P_n^{(ss)}\log P_n^{(ss)}
\ee
using the expression for the steady-state photon number distribution given in
Refs \cite{majumdar,nayak},
 with the normalysation
$\sum_n P_n = 1$. It is straightforward to derive the expression for the
 density operator of the cavity field after  passage
of the two experimental atoms\cite{majumdar}. We denote this cavity state
as $\rho_f^{(2)}$, given by
\ben
\rho_f^{(2)} = {\cal A}{\cal A}\rho_f^{(ss)}{\cal A}^{\dg}{\cal A}^{\dg}
+ {\cal D}{\cal D}\rho_f^{(ss)}{\cal D}^{\dg}{\cal D}^{\dg} \nonumber \\
 + {\cal A}{\cal D}\rho_f^{(ss)}{\cal D}^{\dg}{\cal A}^{\dg} +
{\cal D}{\cal A}\rho_f^{(ss)}{\cal A}^{\dg}{\cal D}^{\dg}
\een 
 and then compute its Shannon entropy 
\be
S(\rho_f^{(2)}) = -\sum_{n=1}^{\infty}P_n^{(2)}\log P_n^{(2)}
\ee
as well, through the corresponding photon number distribution
$P_n^{(2)} = \langle n|\rho_f^{(2)}|n\rangle$. 

We observe that
the streaming atoms leave an imprint on the the photon
distribution function, as reflected by its reduced Shannon entropy.
In Fig.2 we also plot the variation of 
$S(\rho_f^{(2)})-S(\rho_f^{(ss)})$ (the difference of the cavity entropy 
before and
after the passage of the atoms)
versus the parameter $D$.  
The difference of the cavity entropy before and
after the passage of the atoms is transported towards constructing the
atomic entanglement. 
A striking feature of our results is the remarkable functional
correspondence between the difference $S(\rho_f^{(2)})-S(\rho_f^{(ss)})$
and the entanglement of formation $E_F(\rho_a)$ for the atoms up to a 
sufficiently long  atom-photon interaction time $t$. The theory of the
micromaser that we are using\cite{majumdar,nayak} takes into account 
dissipation during the  atom-photon interaction time, as well as during 
the much
longer time interval between the passage of the successive atoms.   
Dissipation builds up with the passage of a number of atoms through the
cavity affecting the steady-state statistics\cite{majumdar,nayak}. 
For short atom-photon interaction times before dissipative effects can
creep in, the atomic entanglement exhibits peaks reflective of the 
characteristic micromaser thresholds. Corresponding peaks are observed in
the difference of Shannon entropies characterizing the maximas of information
loss from the cavity around these parameter values. For longer interaction
times (corresponding to $D \ge 30$ in Fig.2) dissipative effects result in
the saturation of the value of $v$ (around $1$) and also of $E_F$. In contrast,
the difference of the Shannon entropies continues to grow signifying
greater loss of information from the cavity to the environment. The role
of dissipative effects on hindering information transfer from the cavity to
the atoms is clearly evident for these parameter values.

\begin{figure}
\begin{center}
\centerline{\epsfig{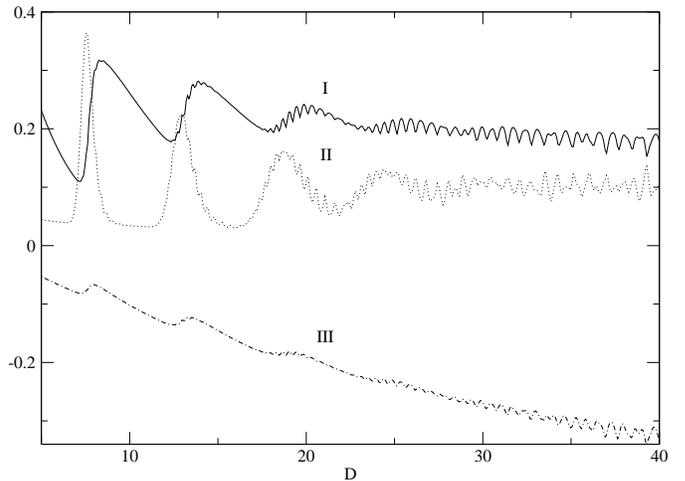}}
\caption {The entanglement of formation $E_F$ (curve I), 
the variance $v/10$ of the photon number distribution (curve II), and
and the loss of Shannon entropy of the cavity, 
$S(\rho_f^{(2)})-S(\rho_f^{(ss)})$ (curve III) are
 plotted with respect 
to parameter $D$.
$N=100$, $\kappa/g = 10^{-6}$, $n_{\mathrm{th}} = 0.01$.}
\end{center}
\end{figure}
To summarize, we have considered a realistic model for the one-atom micromaser
which is feasible to operate experimentally. In this device, correlations
between two separate atoms build up via  atom-photon interactions inside
the cavity, even though
no single atom interacts directly with another. 
The state-of-art of the device\cite{rempe} does not allow dissipative 
effects to reduce drastically the entanglement
generated between a successive pair of atoms for an accessible range of
parameters. Their effects 
nonetheless, are felt
via the photon statistics of the cavity, on the transport of information
from it, more of which is lost to the environment for longer interaction
times. If our analysis 
were to be extended to include the quantification
of the entanglement generated between  three or more
successive atoms, which is an interesting 
possibility within the present framework, then  cavity photon loss 
is  expected to directly
bear upon  such multipartite atomic entanglement. For Rydberg atoms undergoing
microwave transitions, the atomic damping constant $\Gamma=0$. However,
in the one-atom laser operated at optical frequencies, a finite $\Gamma$
will further diminish the degree of entanglement. Thus, the generation
of entangled bipartite atoms using their Rydberg energy levels and a
microwave cavity seems to be more feasible compared to an optical
cavity of the one atom laser.

We have seen that it is possible to demonstrate the transfer of information
through the micromaser set-up. The initial joint state of two successive
atoms that enter the cavity is unentangled. Interactions mediated by the
cavity photon field results in the final two-atom state being of a mixed
entangled type. The information content of the final two-atom state
characterised by its entanglement of formation emanates from the loss of
information of the cavity field quantified by the reduction of Shannon
entropy. Dissipation results in a part of the Shannon entropy, or information
content of the cavity to be lost to the environment. Alternatively, the final 
entangled state of the two atoms could
be viewed as arising through the interaction of two separate atoms with
the common, but suitably tailored ``environment'', the role of which is
played by the cavity field. In this respect, the physics of entanglement
generation in the micromaser is a special case of `environment induced
entanglement'\cite{braun}.  Our approach enables
the study of the relationship between {\it a priori} abstract information 
theoretic measures
for systems in different Hilbert spaces describing the atomic states and
cavity photons respectively. Further studies on dynamics of information 
transfer need to be undertaken in the contexts of different micromaser
models. One particular model is that of the driven micromaser\cite{solano}
which due to its solvable nature seems to be a promising candidate for
investigation of details such as channel capacity of information transfer.
Finally, it would be interesting to
investigate  the recently proposed fascinating
conjecture of Bennett\cite{bennett} regarding the `monogamous' nature of 
entanglement through the micromaser device.

\end{document}